
\newbox\leftpage
\newdimen\fullhsize
\newdimen\hstitle
\newdimen\hsbody
\tolerance=1000\hfuzz=2pt

\def\bigans{b }
\def\answ{b }
\ifx\answ\bigans\message{(this will come out unreduced.}
\magnification=1200\baselineskip=20pt
\font\titlefnt=amr10 scaled\magstep3\global\let\absfnt=\tenrm
\font\titlemfnt=ammi10 scaled\magstep3\global\let\absmfnt=\teni
\font\titlesfnt=amsy10 scaled\magstep3\global\let\abssfnt=\tensy
\hsbody=\hsize \hstitle=\hsize 

\else\def\apans{h }
\message{(this will be reduced.}
\let\lr=l
\magnification=1000\baselineskip=16pt\voffset=-.31truein
\hstitle=8truein\hsbody=4.75truein\vsize=7truein\fullhsize=10truein
\ifx\apansw\apans\special{ps: landscape}\hoffset=-.54truein
  \else\hoffset=.05truein\fi
\font\titlefnt=amr10 scaled\magstep4 \font\absfnt=amr10 scaled\magstep1
\font\titlemfnt=ammi10 scaled \magstep4\font\absmfnt=ammi10 scaled\magstep1
\font\titlesfnt=amsy10 scaled \magstep4\font\abssfnt=amsy10 scaled\magstep1

\output={\ifnum\count0=1 
  \shipout\vbox{\hbox to \fullhsize{\hfill\pagebody\hfill}}\advancepageno
  \else
  \almostshipout{\leftline{\vbox{\pagebody\makefootline}}}\advancepageno
  \fi}
\def\almostshipout#1{\if l\lr \count1=1
      \global\setbox\leftpage=#1 \global\let\lr=r
   \else \count1=2
      \shipout\vbox{\hbox to\fullhsize{\box\leftpage\hfil#1}}
      \global\let\lr=l\fi}
\fi

%

\def\draftmode{\message{ DRAFTMODE }\def\draftdate{{\rm preliminary draft:
\number\month/\number\day/\number\yearltd\ \ \hourmin}}%
\headline={\hfil\draftdate}\writelabels\baselineskip=20pt plus 2pt minus 2pt
 {\count255=\time\divide\count255 by 60 \xdef\hourmin{\number\count255}
  \multiply\count255 by-60\advance\count255 by\time
  \xdef\hourmin{\hourmin:\ifnum\count255<10 0\fi\the\count255}}}

\def\title#1#2{\nopagenumbers\absfnt\hsize=\hstitle\rightline{}%
\centerline{\titlefnt\textfont0=\titlefnt%
\textfont1=\titlemfnt\textfont2=\titlesfnt #1}%
\centerline{\titlefnt\textfont0=\titlefnt%
\textfont1=\titlemfnt\textfont2=\titlesfnt #2
}%
\textfont0=\absfnt\textfont1=\absmfnt\textfont2=\abssfnt\vskip .5in}

\def\date#1{\vfill\leftline{#1}%
\tenrm\textfont0=\tenrm\textfont1=\teni\textfont2=\tensy%
\supereject\global\hsize=\hsbody%
\footline={\hss\tenrm\folio\hss}}
%

\def\nolabels{\def\eqnlabel##1{}\def\eqlabel##1{}\def\reflabel##1{}}
\def\writelabels{\def\eqnlabel##1{\hfill\rlap{\hskip.09in\string##1}}%
\def\eqlabel##1{\rlap{\hskip.09in\string##1}}%
\def\reflabel##1{\noexpand\llap{\string\string\string##1\hskip.31in}}}
\nolabels
%
\global\newcount\secno \global\secno=0
\global\newcount\meqno \global\meqno=1

\def\newsec#1{\global\advance\secno by1\xdef\secsym{\the\secno.}\global\meqno=1
\bigbreak\bigskip
\noindent{\bf\the\secno. #1}\par\nobreak\medskip\nobreak}
\xdef\secsym{}

\def\appendix#1#2{\global\meqno=1\xdef\secsym{#1.}\bigbreak\bigskip
\noindent{\bf Appendix #1. #2}\par\nobreak\medskip\nobreak}


\def\eqnn#1{\xdef #1{(\secsym\the\meqno)}%
\global\advance\meqno by1\eqnlabel#1}
\def\eqna#1{\xdef #1##1{(\secsym\the\meqno##1)}%
\global\advance\meqno by1\eqnlabel{#1$\{\}$}}
\def\eqn#1#2{\xdef #1{(\secsym\the\meqno)}\global\advance\meqno by1%
$$#2\eqno#1\eqlabel#1$$}

\global\newcount\ftno \global\ftno=1
\def\refsymbol{\ifcase\ftno
\or\dagger\or\ddagger\or\P\or\S\or\#\or @\or\ast\or\$\or\flat\or\natural
\or\sharp\or\forall
\or\oplus\or\ominus\or\otimes\or\oslash\or\amalg\or\diamond\or\triangle
\or a\or b \or c\or d\or e\or f\or g\or h\or i\or i\or j\or k\or l
\or m\or n\or p\or q\or s\or t\or u\or v\or w\or x \or y\or z\fi}
\def\foot#1{{\baselineskip=14pt\footnote{$^{\refsymbol}$}{#1}}\ %
\global\advance\ftno by1}


\global\newcount\refno \global\refno=1
\newwrite\rfile
\def\ref#1#2{$^{(\the\refno)}$\nref#1{#2}}
\def\nref#1#2{\xdef#1{$^{(\the\refno)}$}%
\ifnum\refno=1\immediate\openout\rfile=refs.tmp\fi%
\immediate\write\rfile{\noexpand\item{\the\refno.\ }\reflabel{#1}#2.}%
\global\advance\refno by1}
\def\addref#1{\immediate\write\rfile{\noexpand\item{}#1}}

\def\semi{;\hfil\noexpand\break}



\hyphenation{anom-aly anom-alies coun-ter-term coun-ter-terms}


\def\cL{\hbox{{$\cal L$}}}
 


   
  \def\bi{{\bf i}}

\def\pmb#1{\setbox0=\hbox{#1}%
 \kern-.025em\copy0\kern-\wd0
 \kern .05em\copy0\kern-\wd0
 \kern-.025em\raise.0433em\box0 }


\def\cL{{\cal L}}
\def\cP{{\cal P}}
\def\cK{{\cal K}}
\def\cZ{{\cal Z}}
\def\bR{{\bf R}}

\centerline{{\bf Formulating a first-principles statistical theory of}}
\centerline{{\bf growing surfaces in two-dimensional Laplacian fields}}

\bigskip
\centerline{{\bf Raphael Blumenfeld}}

\centerline{Center for Nonlinear studies and the Theoretical Division, MS B258}
\centerline{Los Alamos National Laboratory, Los Alamos, NM 87545, USA}

\bigskip
\item{}{\bf Abstract}
\smallskip
A statistical theory of two-dimensional Laplacian growths is formulated from
first-principles. First the area enclosed by the growing surface is mapped
conformally to the interior of the unit circle, generating a set of dynamically
evolving quasi-particles. Then it is shown that the evolution of a
surface-tension-free growing surface is Hamiltonian. The Hamiltonian
formulation allows a natural extension of the formalism to growths with either
isotropic or anisotropic surface tension. It is shown that the curvature term
can be included as a surface energy in the Hamiltonian that gives rise to
repulsion between the quasi-particles and the surface. This repulsion prevents
cusp singularities from forming along the surface at any finite time and
regularizes the growth. An explicit example is computed to demonstrate the
regularizing effect. Noise is then introduced as in traditional statistical
mechanical formalism and a measure is defined that allows analysis of the
spatial distribution of the quasi-particles. Finally, a relation is derived
between this distribution and the growth probability along the growing surface.
Since the spatial distribution of quasi-particles flows to a stable limiting
form, this immediately translates into predictability of the asymptotic
morphology of the surface. An exactly-solvable class of arbitrary initial
conditions is analysed explicitly.


\date{LA-UR-93-4338}
\eject

\newsec{Introduction}
\bigskip

Growing surfaces in diffusion-controlled, and generally Laplacian fields have
been the focus of much attention recently. These processes are very simple to
formulate but extremely difficult to analyse theoretically. Many examples of
such growths can be found where the resulting morphologies are very ramified
and generally exhibit a rich variety of patterns. Well-known cases are
dendritic growth and solidification in supercooled liquid, diffusion-limited
aggregation, electrodeposition, viscous fingering in Hele-Shaw cells, growth of
bacterial colonies on an agar substrate, and many more.\ref\rev{For review see,
e.g., P. Pelce, "Dynamics of Curved Fronts" (Academic Press, San Diego, 1988);
D. A. Kessler, J. Koplik and H. Levine, Adv. Phys. {\bf 37}, 255 (1988); P.
Meakin, in "Phase Transitions and Critical Phenomena" Vol. 12 (Academic Press,
1988) Eds. C. Domb and J. L. Lebowitz; T. Vicsek, "Fractal Growth Phenomena"
(World Scientific, Singapore, 1989)} The inherent difficulty in understanding
these processes analytically stems from the characteristic instability of the
moving boundary, combined with screening competition of the growing arms over
the field. Consequently the number of theoretical predictions that relate to
such processes is surprisingly small compared to the large amount of existing
phenomenological and numerical data.

There currently exist several variants of a renormalization group
approach,\ref\rgi{H. Gould, F. Family and H. E. Stanley, Phys. Rev. Lett. {\bf
50}, 686 (1983); T. Nagatani, Phys. Rev. {\bf A 36}, 5812 (1987); J. Phys. {\bf
A 20}, L381 (1987); X. R. Wang, Y. Shapir and M. Rubinstein, Phys. Rev. {\bf A
39}, 5974 (1989); J. Phys. {\bf A 22}, L507 (1989); P. Barker and R. C. Ball,
Phys. Rev. {\bf A 42}, 6289 (1990)} and more recently two such generic
approaches managed to yield rather accurate values for the scaling of the
radius of gyration of several growths.\ref\piet{L. Piteronero, A. Erzan and C.
Evertsz, Phys. Rev. Lett. {\bf 61}, 861 (1988), Physica {\bf A151}, 207
(1988)}$^,$\ref\tch{T. C. Halsey and M. Leibig, Phys. Rev. {\bf A 46}, 7793
(1992); } Generically, in such approaches an iterative procedure is carried out
for the growth probability density. These approaches assume existence of a
limiting stable distribution with {\it scaling} properties and analyse the
scaling exponent of the average mass with time or scale. However, a full theory
that starts from the fundamental Eqs. of motion (EOM) and leads in a
step-by-step manner to predicting the full asymptotic structure of the surface,
without those assumptions, is yet to be put forward.

Mostly due to conceptual and technical simplicity most of the literature treats
growth of such patterns in two-dimensions, and this discussion is no exception.
I focus here on this case mainly because it allows for an elegant conformal
formulation, which simplifies the formalism. Nevertheless, it should be
stressed that the essential features that are relevant to the present theory
can be applied to growth processes and evolving surfaces in higher dimensions,
as will be shown elsewhere.\ref\bi{R. Blumenfeld, in preparation}

Already in the forties\ref\galin{L. A. Galin, Dokl. Akad. Nauk USSR {\bf 47},
246 (1945); P. Ya. Polubarinova-Kochina, Dokl. Akad. Nauk USSR {\bf 47}, 254
(1945); Prikl. Math. Mech. {\bf 9}, 79 (1945)} and more recently\ref\sb{B.
Shraiman and D. Bensimon, Phys. Rev. {\bf A 30}, 2840 (1984)} it has been
proposed that in the case of two dimensional growth of a surface-tension-free
boundary in a Laplacian field, conformal mapping can be used to transform the
problem to the dynamics of a many-body system. This approach converts the
problem of solving a one parameter-dependent PDE to that of solving a system of
first order ODE's, with each ODE corresponding to an EOM of one quasi-particle
(QP) of the equivalent many-body system. This set of ODE's turns out to be
strongly coupled and nonlinear, making the problem still very difficult to
solve, other than in special cases.\sb$^,$\ref\all{L. Paterson, J. Fluid Mech.
{\bf 113}, 513 (1981); L. Paterson, Phys. Fluids {\bf 28}, 26 (1985); S. D.
Howison,  J. Fluid Mech. {\bf 167}, 439 (1986); D. Bensimon and P. Pelce, Phys.
Rev. {\bf A33}, 4477 (1986); S. Sarkar and M. Jensen, Phys. Rev {\bf A35}, 1877
(1987); B. Derrida and V. Hakim, Phys. Rev. {\bf A45}, 8759 (1992)}

Thus this technique has found very little use in the research community.
Moreover, this elegant description suffers from an even more acute problem. In
most cases (i.e., for most initial conditions) the EOM break down after a
finite time due to the inherent instability (of the Mullins-Sekerka
type\ref\ms{W. W. Mullins and R. F. Sekerka, J. Appl. Phys. {\bf 34}, 323
(1963)}) of the surface with respect to growth of perturbations on arbitrarily
short lengthscales. In the absence of surface tension, these develop into cusp
singularities along the physical surface, which correspond in the mathematical
plane to zeroes or poles of the conformal map arriving at the UC at a finite
time. There have been some attempts to suppress this catastrophe by adding
surface tension and using it to cut off the short lengthscales in a
renormalizable manner.\ref\leo{D. Bensimon, L. P. Kadanoff, S. Liang, B. I.
Shraiman and C. Tang, Rev. Mod. Phys. {\bf 58}, 977 (1986); W-s Dai, L. P.
Kadanoff and S. Zhou, Phys. Rev. {\bf A 43}, 6672 (1991)} These approaches,
however, seem to be somewhat ad hoc in the sense that the procedure for cutting
off the short scales can be arbitrarily chosen. In other words, one can
introduce a phenomenological renormalizing procedure for the
surface-tension-dependent term in the EOM of the physical surface, which
readjusts the EOM of the singularities of the map and prevents the breakdown.
The choice of the phenomenological term is arbitrary in a sense and
renormalization approaches are known to introduce uncontrolled errors. A
perturbative approach is also difficult because an arbitrarily small surface
tension turns out to be a singular perturbation for the unregularized system,
which, for a small surface tension, makes the sutface's evolution very
sensitive to initial conditions.\ref\tanveer{S. Tanveer, Philosophical
Transactions of The Royal Society (London), {\bf A 343}, 155 (1993) and
references therein} Another approach that has been proposed to prevent
formation of such cusps relies on a recent observation\ref\bbi{R. Blumenfeld
and R. C. Ball, in preparation} that tip-splitting reduces local values of high
surface curvature energy. It was therefore proposed that the mathematical QP
split when they come too close to the surface, implying a field theoretical
description of the problem.

Another issue in this context which is significant for the purpose of the
formalism presented here is whether the system can be described by a
Hamiltonian structure.  It has been long known that this problem enjoys a set
of conserved quantities,\ref\rich{S. Richardson, J. Fluid Mech., {\bf 56}, 609
(1972)}$^,$\ref\mm{M. B. Mineev, Physica {\bf D 43}, 288 (1990)} but the
usefulness of these for integrating the EOM was not clear. This issue has been
recently addressed by this author\ref\bii{R. Blumenfeld, Phys. Lett. {\bf A},
To appear (1994)} and it appears that indeed the system enjoys a Hamiltonian
formulation, which under a given condition can even be integrated, as has been
demonstrated for a specific family of initial conditions.

In this paper I try to lay the foundations for a full theory of growth of such
surfaces. The theory is constructed in five stages: 1) First the EOM of the
surface, which is generally a first order PDE, is converted into a set of first
order ODE's for quasi-particles of a many-body system, as mentioned above. 2) A
Hamiltonian structure of the system is formulated. 3) Taking advantage of the
existence of a Hamiltonian functional I introduce the surface contribution as
simply another energy term in the Hamiltonian. This term prevents occurrence of
cusp singularities and makes the formulation valid for all times and for any
initial condition. In the many-body Hamiltonian system this term gives rise to
an effective {\it repulsion} between the surface and the QP. This approach is
suited to anisotropic, as well as isotropic surface tension. 4) Next the master
equation that governs the evolution of the spatial distribution of the QP is
formulated. This distribution flows to a stable limiting form. At this stage
noise is naturally introduced into the theory in a fashion similar to
traditional statistical mechanical theories. 5) The last step consists of
translating the spatial distribution of the QP into the statistics of the
growing surface, thus enabling to analyse and predict from {\it
first-principles} the morphological features of the asymptotic pattern.

\smallskip
\newsec{Formulation of the problem and mapping into many-body dynamics}
\bigskip

The two-dimensional problem under study can be formulated as follows: Consider
a simply connected line surface, $\gamma(s)$, parametrized by an angular
variable, $0\leq s <2\pi$. On this boundary the potential field, $\Phi$, is
fixed at a given value, $\Phi_0$. This field can represent an electrostratic
potential, a concentration field for diffusion controlled processes, a thermal
field, and much more. A higher potential value is assigned to a circular
boundary of radius $R_\infty$ that is much larger than the size of the area,
$S$, enclosed by $\gamma$. Assuming no sources, the field outside $S$, $\Phi$,
is determined by Laplace's Eq.
\eqn\L{\nabla^2 \Phi = 0\ .}
The surface is assumed to grow according to a constitutive rule that relates
the local rate of growth proportionally to the local gradient of the field
normal to the surface
$$v_n = - {\bf \nabla}\Phi\cdot\hat{\bf n}\ .$$
Being two-dimensional this process allows for an elegant use of complex
analysis to write down a closed form evolution equation.\galin$^,$\sb$^,$\rich
First one maps the closed curve, $\gamma$, in the $\zeta$ complex plane
conformally to the unit circle (UC) in a mathematical $z$-plane:
$$\zeta = F(z)\ .$$
In the $z$-plane the complex potential field is simply
$$\Phi(z) = \ln{(z/z_0)} = \ln{(|z|/|z_0|)} + i[{\rm arg}(z)-{\rm arg}(z_0)]\
,$$
where $z_0$ is some constant that is determined by the reference potential on
the UC. So the complex field $\nabla\Phi$ along the physical surface is
\eqn\Yi{-\nabla\Phi(\zeta) = -\Biggl({{\partial\Phi(\zeta)}\over
{\partial\zeta}}\Biggr)^* = {{-i}\over{(z F')^*}}\ ,}
where $^*$ stands for complex conjugate and the prime indicates derivative with
respect to $z$. For the moving surface $z\to e^{is}$ ($0\leq s < 2\pi$) and
therefore the actual curve evolves according to
\eqn\Yii{{{\partial\gamma(s,t)}\over{\partial t}} = -i
\Biggl({{\partial\gamma^*(s,t)}\over{\partial s}}\Biggr)^{-1}\ . }
Since this equation is obtained by monitoring only the normal velocity of the
boundary at each $s$, it does not maintain the right parametrization with time.
To correct this one has to allow for a tangential velocity, so that a point can
'slide' along the curve. This is essential for the purpose of studying the
evolution of the surface through the dynamics of the map $F$. For the map to be
conformal both $F$ and its inverse must be analytic in $z$ outside the UC. But
maintaining only the normal velocity spoils this analyticity because the
components of the gradient of the potential field are not analytic. Therefore,
if $\gamma(s,t)$ is to be described as the limit
$$\gamma(s,t) = \lim_{z\to e^{is}} F(z,t)\ ,$$
then the r.h.s. of $\Yii$ needs to be augmented. The augmented evolution
equation for the surface reads\sb
\eqn\Ai{\partial_t\gamma(s,t) = -i\partial_s\gamma(s,t)
\Bigl[|\partial_s\gamma(s,t)|^{-2} +
ig\Bigl(|\partial_s\gamma(s,t)|^{-2}\Bigr)\Bigr]\ .}
The first term on the r.h.s. represents the field gradient normal to the
surface as before. The second term that is obtained through the demand that the
r.h.s. is analytic in $z$, and represents the tangential component of the
velocity, which causes the 'sliding' of a point $s$ along $\gamma$. Although
mathematically important, this component has no real physical consequence for
the advance of the surface since we can reparametrize the curve as we wish at
each time step.

One can write now the EOM for the map $F$, which gives $\Ai$, but for the
purpose of this presentation it is more convenient to write the EOM of
$F'(z,t)\equiv dF(z,t)/dz$:
\eqn\Aia{\partial F'/\partial t = {\partial\over{\partial z}} \bigl\{z F'
G\bigr\}\ . }
The form of $F'$ considered here is chosen to generally consist of a ratio of
two polynomials.  It turns out that one of the constraints on the
map\bbi$^,$\bii is that these polynomials are of the same degree so that the
map should preserve the topology far away from the growth. Namely, that
$\lim_{z\to\infty}F(z,t)=Az$, where $A$ is a space-independent global scaling
prefactor. This requirement amounts to leaving the original boundary conditions
at $R_\infty$ unchanged. The map now is
\eqn\Aii{ F'(z,t) = A(t)\prod_{n=1}^N {{z-Z_n}\over{z-P_n}}\ .}
The quantities $\{Z_n\}$ and $\{P_n\}$ represent the locations of the zeros and
the poles of the map, respectively. As discussed below, the number of each
species, $N$, may actually go to infinity by treating the local densities of
these species, but for the sake of clarity I will discuss in this presentation
only discrete cases. Since the map and its inverse are conformal then these
poles and zeros must be confined to the interior of the UC that is mapped to
the interior of the growth. Thus the evolution of the surface can be now
expressed in term of the dynamics of these zeros and poles. By inserting the
explicit form of $F'$ into $\Aia$, rearranging terms, and then comparing the
residues on both sides of the resultant equation, we arrive at a set of first
order ODE's for the location of the zeros and the poles:\sb$^,$\bbi
\eqn\Aiii{\eqalign{-\dot Z_n &= Z_n\Bigl\{G_0 + \sum_{m'} {{Q_n +
Q_{m'}}\over{Z_n - Z_{m'}}} \Bigr\} + Q_n\Bigl\{1 - \sum_m{{Z_n}\over{Z_n -
P_m}}\Bigr\} \equiv f^{(Z)}_n(\{Z\};\{P\}) \cr
-\dot P_n &= P_n\Bigl\{G_0 + \sum_m {{Q_m}\over{P_n - Z_m}}\Bigr\} \equiv
f^{(P)}_n(\{Z\};\{P\})\ , \cr }}
where
$$\eqalign{Q_n &= 2\prod_{m=1}^N {{(1/Z_n - P_m^*)(Z_n-P_m)}\over
{(1/Z_n - Z_m^*)(Z_n - Z_{m'})}} \cr
G_0 &= \sum_{m=1}^N {{Q_m}\over{2 Z_m}} + \prod_{m=1}^N {{P_m}\over{Z_m}}\ ,
\cr}$$
and where I have adopted the convention that the primed index indicates $m'\neq
n$. From the equations $\Aia$, $\Aii$ and $\Aiii$, one can construct the
evolution equation for $\ln A(t)$. In the following I will disregard the
evolution of this scale factor, which is unimportant to the main thrust of this
presentation. The implication of this is that at each time step the growth is
in fact rescaled such that the prefactor is always one. The locations of these
zeros and poles can be now considered as QP of a many-body system that follow
the trajectories of Eq. $\Aiii$.

Thus the problem of the growing surface has been transformed into the problem
of the analysis of a many body system. These general EOM have been analysed in
various limits\bbi and for several particular initial conditions.\all$^,$\leo
Such an analysis is not the purpose here. Rather, since the present aim is at a
general theory I now turn directly to formulate the Hamiltonian of the system.

\bigskip
\newsec{The Hamiltonian formulation}
\smallskip

The Hamilton form into which we wish to map the system
$$H=H(\{J\};\{\Theta\})\ ,$$
is generally a function of new canonical complex variables that depend on the
coordinates $\{Z\}$ and $\{P\}$. First, let me convince the reader that there
is ground for a belief in such a formulation. The first hint can be found in
the EOM of the surface $\Yii$. This equation can be slightly changed to read
\eqn\Yiii{{{\partial\gamma(s,t)}\over{\partial t}} = -i {{\delta
s}\over{\delta\gamma^*(s,t)}}\ , }
where the spatial partial derivative has been replaced by the $\delta$
operator. This form, however, represents exactly Hamiltonian description if
$\gamma$ is interpreted as a field (complex) variable and $s$ plays the role of
an energy functional of $\gamma$.\ref\s{M. Mineev-Weinstein, Private
communication} Thus it appears that a Hamiltonian formulation does underlie the
physical process. With this realization it is tempting to ask whether the
addition of the tangential velocity makes a difference to this conclusion. To
show that this is not so consider the EOM $\Aia$. This equation can be
interpreted as one of Hamilton's equations, where $\tilde H = z F' G$ is a
Hamiltonian density and $F'$ is one of the conjugate variables. Using the
identity
\eqn\Aiva{\Biggl({{\partial F'}\over{\partial t}}\Biggr) \Biggl({{\partial
t}\over{\partial z}}\Biggr) \Biggl({{\partial z}\over{\partial F'}}\Biggr) =
-1}
and combining with Eq. $\Aia$ immediately yields the relation
\eqn\Aivb{{{\partial z}\over{\partial t}} = -{{\partial (z F' G)}\over{\partial
F'}}\ .}
Again we have arrived at a Hamiltonian description: Eqs. $\Aia$ and $\Aivb$
constitute the conjugate pair of Hamilton's relations if $z$ is interpreted as
the variable that is formally conjugate to $F'$.

Translating the Hamiltonian back to the physical $\zeta$-plane, $\tilde H(z) =
H(\zeta)$, and rewriting the EOM for the surface in the original coordinates we
obtain
\eqn\Aivc{{\partial\over{\partial \zeta}}H(\zeta) = {\partial\over{\partial t}}
\ln F' = \sum_{n=1}^N {{\dot P_n}\over{F^{-1}(\zeta) - P_n}} - {{\dot
Z_n}\over{F^{-1}(\zeta) - Z_n}}\ .}
Inspecting the EOM $\Aiii$ immediately reveals that those are the exact direct
consequence of $\Aivc$ when a contour integral over $\partial H /
\partial\zeta$ is taken around a close neighbourhood of the location of the QP
in the $\zeta$-plane. Hence it is the contour integration of $H$ that connects
surface dynamics to the many-body formulation. Therefore, since there is a
Hamiltonian description that underlies the surface dynamics it makes sense that
the system of the $2N$ QP (i.e., of $4N$ degrees of freedom) is Hamiltonian
too. The construction of a Hamiltonian directly to the many-body system has
been recently carried out,\bii where it has been shown that under a given
condition the Hamiltonian is even integrable. Indeed such integrability has
been recently demonstrated explicitly for a class of arbitrary initial
conditions. This class is generalized in the Appendix.

It should be remarked at this point that the above discussion suggests that the
formulation presented here can be generalized to a {\it continuous} density of
zeroa and poles as follows: Inspection of the EOM and the signs of the residues
in Eq. $\Aia$ shows that we can interpret the zeros and poles as positive and
negative charges, respectively. Then the contour integrals in the plane, that
relate the map's evolution to the EOM of the charges can be regarded as Gauss
integration around an area that contains some distribution of charges. Since
all we know is the value of such the integral over $\partial H / \partial\zeta$
we can attribute it to a continuous, rather than to a discrete, density of
charges. The result will be a first order ODE for the {\it charge density} in
this region. So by making the typical number of singularities within such an
area very large we effectively pass to the continuum limit. With this extension
of the formalism one can overcome quite a few difficulties that stem from the
finiteness of the number of the singularities\ref\commi{When the number of
singularities is finite, the EOM, as given in Eq. $\Aiii$ preserve this number.
This is unphysical in many systems that undergo side-branching and
tip-splitting, because such processes typically correspond to production and
annihilation of singularities of the map, as discussed by Blumenfeld and
Ball.\bbi The extension to a continuous density of such singularities and field
formulation makes it possible to overcome this problem} and generally pass to a
continuous field description of the problem.

\smallskip
\newsec{Introduction of surface tension}
\bigskip

So far I have considered a free boundary (i.e. without surface-tension) that
evolves in an external Laplacian field. I now turn to discuss the effect that
surface tension has on the growing process. Evidently, patterns that result
from real growth processes do not entertain any cusps forming along the
boundary. Depending on the system, this is either due to some microscopic
atomistic cutoff scale, below which the above description ceases to apply, or
because there is a macroscopic surface energy to be paid when the curvature of
the surface increases. Here I focus on the second mechanism for two reasons: i)
The entire formulation presented here relies on the continuous aspect of the
surface and therefore makes it cumbersome to treat atomistic cutoffs; ii) Many
natural growth processes can be shown to enjoy a continuum description where
the surface energy can be defined as a function of a continuous curvature,
which is bounded along the surface.

Assuming then that there is a given surface energy that has a smoothening
effect on the boundary, the question is what would this effect translate into
in the context of the many-body system? To answer this question one should
first note that the radius of curvature that the QP enhances along the surface
increases with the distance between a QP and the boundary, namely, the closer
is the QP to the boundary, the sharper the distortion of the surface. The sign
of this effect depends directly on the 'charge' of the QP with protrusions
corresponding to zeros and indentations to poles. For example, for a relatively
isolated zero at
$$Z_0 = (1-\delta)e^{is_0} \hskip1cm ; \hskip1cm \delta \ll 1$$
the curvature at $s_0$, $K(s_0)\sim 1/\delta^3$ (see next section). Therefore,
the effect of surface energy to prevent too small radii of curvature should be
translated into preventing QP from coming too close to the boundary. From this
argument it follows that in the equivalent many-body Hamiltonian system the
effect of a positive surface tension must correspond to {\it repulsion} between
the QP and the inside of the boundary. It needs to be emphasized here that only
because we have a Hamiltonian formulation available, the term 'repulsion' can
be used with any physical meaning. The Hamiltonian structure makes it possible
to account for such effects in a natural energetic context, while without it
surface effects could only be incorporated by introducing an additional ad-hoc
term into the EOM.

To completely prevent cusps the repulsive potential between the surface and the
QP must {\it diverge} as their separation vanishes. It follows that in this
case we can consider the QP to be effectively confined to within a potential
well that consists of an infinite wall (the surface boundary).
\smallskip

{\bf Example}:

To demonstrate how this method regularizes the growth let us consider the
simple case discussed in the Appendix. The initial conditions of this growth
process consist of $N$ pairs of zeros and poles arranged symmetrically on $N$
rays. The EOM's for this system, (A.3), need to be augmented with the surface
potential. The choice of the model potential is at our disposal at this stage
(see a more detailed discussion on the surface potential below). For the
purpose of the present example let us assume a form that gives rise to an
arbitrary (negative) power, $\alpha$, of the distance between the QP and the
UC. The full EOM's now become
\eqn\Di{\eqalign{ -{1\over {N}} {\dot x \over x} &= y/x - \cK \bigl[1 - 2/N +
(1 + 2/N)y/x \bigr] + \sigma_x / (1-x)^\alpha \cr
-{1\over {N}} {\dot y \over y} &= y/x - \cK (1 + y/x) + \sigma_y /(1-y)^\alpha
\cr}}
where $\sigma_x$, $\sigma_y$ are constants and the other notations are as in
Eq. (A.3). An analysis of these equations shows that the growth becomes uniform
very quickly. In fig. 1 I plot the resulting surface with, and without, the
surface term for $\alpha=1$ and for 3 pairs of QP. Without the regularizing
terms fig. 1a shows the formation of cusp singularities at $t=0.2673$
(arbitrary units). With the surface terms the growth is observed to settle into
a uniform process even at times that are orders of magnitude larger. To
demonstrate the uniformity of growth I rescale the area enclosed by the surface
at each time step by $A(t)$ whereupon it can be seen that, asymptotically, the
curves at different times collapse on top of each other. These processes has
been run for times up to $t=1000$ in these arbitrary units to check that the
asymptotic form is stable and does not change. Various different surface
potentials have been found to produce very similar results. A general analysis
of the dependence of the growth on the form of the functional properties of the
surface potential is not intended here and is a subject for future research.

Thus with this regularization the above formalism of dynamics of singularities
has been practically extended to hold up for $t\to\infty$, which has been
heretofore one of the main disadvantages of this general approach. The only
part that still needs to be clarified is whether the form of the map can still
be described by a ratio of polynomials once the Hamiltonian is augmented by the
surface term. It is this author's belief that this is so, albeit with the
possibility of encountering time-dependent, or even infinite, number of
singularities, depending on the form of the repulsive potential. A related
approach has been considered recently by Blumenfeld and Ball,\bbi where a
surface term was introduced in the EOM. Only in that case the proximity to the
surface initiated production of opposite charges (poles and zeros in pairs)
that acted to reduce the local curvature in front of an approaching zero. I
will just remark here that that approach opens the door to a general
interpretation of the field that is induced by the boundary and which is felt
by the QP: In the presence of this field particles can be spontaneously created
by vacuum fluctuations, and annihilated by encountering anti-particles. This
interpretation fits quite naturally in the present formalism because a pole and
a zero do indeed annihilate upon encounter as mentioned already.

Let us now turn to discuss the form of the potential term in more detail. This
term must be a functional of the curvature $K$, which, in turn, can be
expressed in terms of the locations of the QP in the $z$-plane\bbi
\eqn\Dii{K\bigl(s,\{Z\},\{P\}\bigr) = \lim_{z\to e^{is}} \mid F'\mid^{-1}
\Bigl\{ 1 + {\rm Re}\sum_{n=1}^N\bigl\{ {z\over{z-Z_n}} - {z\over{z-P_n}}
\bigr\} \Bigr\}\ .}
So if the Hamiltonian of the surface-free system was $H_0$ then the new
Hamiltonian becomes
\eqn\Diiia{H_K = H_0 + V\Bigl(K\bigl(z,\{Z\},\{P\}\bigr)\Bigr)\ ,}
and the new EOM in the mathematical plane are derived from the new Hamiltonian
as before:
\eqn\Diiib{\partial F'/\partial t = {\partial\over{\partial z}} \left\{z F' G +
V\bigl(K\bigl(z,\{Z\},\{P\}\bigr)\right\}\ .}
Since the potential $V\bigl(K\bigl(z,\{Z\},\{P\}\bigr)$ should effect repulsion
between the QP and the boundary, then the sign of $V$ is immediately
determined.
For example, the simplest form that comes to mind for such a repulsive
potential in the {\it mathematical plane} is
\eqn\Diii{{\rm Re}\left\{V\Bigl(K\bigl(z,\{Z\},\{P\}\bigr)\Bigr)\right\} =
\sigma(z) \Bigl[\mid F'\mid K\bigl(z,\{Z\},\{P\}\bigr)\Bigr]\ .}
The complex form of $V$ can be found from the demand that this term is analytic
outside the growth, which leads to
\eqn\Diiii{V = {1\over{2 \pi i}}\lim_{\epsilon\to 0} \oint {{z + z'}\over{z +
\epsilon - z'}} {\rm Re}\left\{V\right\}
{{d z'}\over{z'}}\ .}
It is important to note that the surface tension $\sigma$ in this formulation
can depend on $z$, therefore allowing for {\it anisotropic surface effects},
e.g., as in crystal growth. The reason for taking the potential term in the
mathematical, rather than in the physical, plane is that it is there that the
QP are moving and where they feel the effects of the 'wall' along the UC. Note
also that The form in $\Diii$ is easy to handle because it decouples naturally
to a sum of individual contributions of the QP:
\eqn\Div{\eqalign{{\rm Re}\left\{V_0\right\} &= \sigma(z) \cr
{\rm Re}\left\{V(Z_n)\right\} &= \sigma(z) {\rm Re} {z\over{z-Z_n}} \cr
{\rm Re}\left\{V(P_n)\right\} &= -\sigma(z) {\rm Re} {z\over{z-P_n}}\ . \cr}}

\smallskip
\newsec{Effects of noise and a statistical formulation of the theory}
\bigskip

The next, and probably technically the most difficult, step towards a theory of
growth involves including the effect of noise. As is well known in growths
governed by Laplacian fields, the patterns that such processes evolve into
depend in a crucial way on the characteristics of the noise in the system. This
noise can originate from many sources: general fluctuations in the local
Laplacian field, discretization of the underlying background over which the
field is solved (lattice growth), discretization of the incoming flux in the
form of finite size particles that stick to the growing aggregate (e.g.,
off-lattice diffusion-limited-aggregation, electrodeposition and similar
processes), etc.. The noise can be also generally correlated in space and in
time. In the present formulation one way to incorporate all these effects is to
interpret them as simply 'smearing' the noiseless predetermined trajectories of
the QP in the mathematical plane. This interpretation enjoys a new meaning now
that we have a Hamiltonian available: The existence of a Hamiltonian
immediately points to the existence of Liouville's theorem in this system,
namely, that the distribution of the canonical variables in phase space is
incompressible. Thus it is straightforward to write down an EOM for the time
evolution of the distribution of the QP in phase space and consequently it can
be possible to analyse its asymptotic behaviour. This exercise is currently
being carried out by this author and will be reported at a later time. Either
from such a calculation, or via phenomenological argument, one can devise a
measure $\mu\bigl(H(\{Z\},\{P\})\bigr)$, for example, the Gibbs measure,
$e^{-\beta H}$) and calculate {\it average} quantities weighted by this measure
\eqn\Dvi{\langle X \rangle = {1\over \cZ} \int X\
\mu\bigl(H(\{Z\},\{P\})\bigr) d^{N}Z d^{N}P \ ,}
where the partition function $\cZ$ is
$$\cZ \equiv \int \mu\bigl(H(\{Z\},\{P\})\bigr) d^{N}Z d^{N}P\ .$$
Suppose that the Gibbs measure is indeed the relevant measure for this purpose.
Then the Lagrange multiplier, $\beta$, which in traditional statistical
mechanics is associated with the temperature, would correspond here to the
effective magnitude of the noise. This issue is also under current
investigation. I should only comment that this approach should turn out to be
equivalent to introducing noise directly in the EOM $\Aiii$, and that such an
equivalence should be possible to elucidate via an argument analogous to the
fluctuation-dissipation theorem.

This formalism gives a well defined framework to describe the statistics of the
QP and in general any property that depends explicitly on the distribution of
their locations. It has been observed time and again that in many growth
processes in Laplacian (and in other) fields the growth probability along the
surface seems to flow towards a stable asymptotic form. One manifestation of
this phenomenon is the appearance of a time-independent multifractal
function.\ref\mf{B. B. Mandelbrot, J. Fluid Mech, ${\bf 62}$, 331 (1974); Ann.
Israel Phys. Soc. ${\bf 2}$, 225 (1978); T. C. Halsey, M. H. Jensen, L. P.
Kadanoff, I. Procaccia and B. I. Shraiman, Phys. Rev. A${\bf 33}$, 1141 (1986);
T. C. Halsey, P. Meakin, and I. Procaccia, Phys. Rev. Lett. ${\bf 56}$, 854
(1986); C. Amitrano, A. Coniglio, and F. diLiberto, Phys. Rev. Lett.
${\bf 57}$, 1016 (1986); P. Meakin, in {\it {Phase \ Transitions \ and \
Critical \ Phenomena}} Vol. ${\bf 12}$, edited by C. Domb and J.L. Lebowitz
(Academic Press, New York 1988), p.335; R. Blumenfeld and A. Aharony, Phys.
Rev. Lett. ${\bf 62}$,  2977 (1989)} Since it is possible to show that the
growth probability along the growing surface is directly related to the spatial
distribution of $\{Z\}$ and $\{P\}$,\ref\biv{R. Blumenfeld, in preparation} one
can therefore analytically predict the statistics of the physical surface, its
{\it asymptotic morphology}, and in particular the entire multifractal
spectrum.

To illustrate how the above formalism is carried out let me outline how to
calculate the asymptotic (steady state) statistics of the curvature along the
surface. It is shown below how knowledge of this distribution yields the
asymptotic growth probability distribution, and how from the latter one can
obtain the fractal dimension and the entire multifractal spectrum. There are
two ways to go about such a calculation:

(I) Since the curvature is a function of the locations of the QP one can simply
calculate the $n$th moment of $K$ from Eq. $\Dvi$ by putting $K^n$ for $X$. For
example, if Gibbs measure can be assumed we have
\eqn\Dvii{M_n(z) \equiv \langle K^n\bigl(z,\{Z\},\{P\}\bigr)\rangle =
{1\over{\cZ}}
\int K^n\bigl(z,\{Z\},\{P\}\bigr) e^{-\beta(H_0 + V)} d^{N}Z d^{N}P\ .}
This integral depends on $z$ because the curvature is a local quantity and in
fact we are probing the statistics at a given location $z=e^{is}$. For
isotropic growths one can integrate the result over $s$ (along the surface) for
the final answer, but for anisotropic growths ($\sigma=\sigma(z)$) expression
$\Dvii$ shows that the curvature statistics may well depend on the direction,
which should not come as a big surprise.

(II) The second approach is to first construct the master equation for the
distribution of the locations of the QP, $\cP$, by using Liouvlle's theorem:
\eqn\Dviii{{{\partial \cP}\over{\partial t}} + \sum_n f_n^{(R_n)} {{\partial
\cP}\over{\partial R_n}} = 0\ ,}
where $R_n$ is the $n$th component of the 4N-dimensional (2N degrees of freedom
in two dimensions) vector ${\bf R} = (Z_1,\dots,Z_N,P_1,\dots,P_N)$. Since all
exisiting observations of Laplacian growth processes indicate that the surface
flows into a well-defined asymptotic morphological form with well-charted
statistics there must exist a stable limiting form that corresponds to the
surface's statistics. Thus one gets a well-defined fractal dimension and a
reproducible multifractal spectrum. We can therefore assume that there is a
nontrivial steady-state solution where the direct dependence of $\cP$ on $t$
vanishes, which simplifies Eq. $\Dviii$. Upon solution of this equation
(clearly under some physically valid approximations, as usually done in
statistical mechanics) one obtains $\cP(\{Z\},\{P\})$. Since the value of the
local curvature $K$ depends on the locations vector $\bR$ one can consequently
find the distribution function of $K$.

\smallskip
{\bf Example}:

Let us consider one possible approximation: Suppose that a QP, indexed 0, is
located at $R_0 e^{is_0}$ where $R_0$ is close to one, and suppose that there
is no other QP closer to the UC in the vicinity of $s_0$. This particular QP
then dominates the local curvature at $z=e^{is_0}$ as can be seen from
expression $\Dii$. Therefore, the approximation consists of completely ignoring
the effect of the other QP at $s_0$. If one further assumes that the
distribution of QP is isotropic then the probability density of $K$,
$\cP_1(K)$, is simply defined in terms of the probability density of $R_0$,
$\cP_0(R_0)$, as follows:
\eqn\Dviv{\cP_1(K) = \cP_0(R_0) \left|{{d R_0}\over{d K}}\right|\ .}
With the above approximation the curvature at $s_0$ becomes
$$K \approx C {{1 + R_0}\over{(1 - R_0)^2}} \approx {{2 C}\over{(1 - R_0)^2}}\
,$$
where $C=\left|\prod_n(e^{is_0}-P_n)/\prod_{m'\neq 0}(e^{is_0}-Z_{m'})\right|$
is approximately constant for $z$ in the neighbourhood of $s_0$.
Differentiating this expression with respect to $R_0$, inserting in $\Dviv$ and
expressing $R_0$ in terms of $K$ gives
\eqn\Dvv{\cP_1(K) = Const. \left({{2 C}\over K}\right)^{3/2} \cP_0\left[R_0 = 1
- \left({{2 C}\over K}\right)^{1/2}\right]\ .}

This alternative route gives again an anisotropic $z$-dependent probability
density of $K$ as discussed above.

A significant point to note in this simple calculation is that even if $\cP_0$
possesses exponentially decaying tails the probabily density of $K$ decays {\it
algebraically}. This feature is the signature of scale invariant and fractal
structures. Therefore, already this crude approximation gives us a hint
regarding the origin of the fractality observed in related real growth
processes, such as electrodeposition, diffusion-limited-aggregation,
solidification, Bacterial growth, etc.

It is possible to generalize the above calculation to a multivariate
distribution (i.e., using the general dependence of $K$ on all the $R_n$'s)
without using the above approximation. This is outside the scope of this paper
and will be discussed elsewhere. In any case, the above demonstrates how to use
the knowledge of the distribution of the locations of the QP to determine the
distribution of the curvature. The latter comprises, by definition, the
morphology of the growing surface, which can now be uniquely determined. Thus,
the statistical formulation presented here constitutes the beginning of a full
theory for the problem of a surface growing in a Laplacian field.

\smallskip
\newsec{Discussion and concluding remarks}
\bigskip

To conclude I have formulated here an initial theory for growth of surfaces in
a two-dimensional Laplacian field. This has been carried out in several stages.
First, following previous results, the evolution of the surface has been
transformed into the problem of studying the dynamics of a many-body system.
The quasi-particles (QP) in this system consist of $N$ zeros and $N$ poles of
the conformal map that maps the interior of the growth onto the area enclosed
by the unit circle (UC). In passing I have pointed out how the formalism can be
extended to include an infinite number of QP and subsequently to describe a
continuous density of these particles. The next step was to show that the
growth process is governed by Hamiltonian dynamics and to explicitly write down
this Hamiltonian. Whether the Hamiltonian is integrable or not is not directly
relevant to the theory formulated here, but a general class of arbitrary
initial conditions that result in integrable dynamics, has been explicitly
analysed and solved for in the appendix. The existence of a Hamiltonian that
underlies the dynamics immediately opens new horizons which I exploited for the
construction of the theory. The first difficulty that the Hamiltonian helps to
overcome is the inherent instability of this general growth problem. By
incorporating the surface effect directly in the Hamiltonian as a potential
term, formation of cusp singularities along the surface is eliminated and the
validity of the equations of motion (EOM) is extended to infinite time. I have
argued that the potential term must correspond to a repulsive interaction
between the QP and the surface. Since this potential diverges as a QP
approaches the surface the system is then confined to a well with infinite
walls located on the UC. An explicit example that demonstrates the
regularization by this method has been computed and plotted. This approach is
naturally suited to both isotropic and anisotropic surface tension effects,
which heretofore could only be included in an ad-hoc manner by assuming an
extra term in the EOM of the surface. The present formalism can easily enjoy a
field formulation in the sense that: i) QP can anihilate as can be seen
directly from the form of the map, $\Aii$, and ii) depending on the nature of
the field that the QP move in, they can split. This latter feature has been
suggested and used previously\bbi with the observation that splitting of zeros
is a mechanism that reduces locally high curvatures. In the physical growth
such a split corresponds either to tip-splitting or to side-branching,
depending on the details of the process.

Turning to the statistics of the growth let us first recall what we require
from a full theory of growth. The theory should start from the basic EOM of the
surface and, taking into account the noise that affects the growth process,
should predict statistical properties of the asymptotic morphology that the
surface evolves into. This is based of course on the observations that such an
asymptotic morphology does exist in most Laplacian growths, e.g., in
diffusion-limited aggregation, solidification, electrodeposition, bacterial
growth and other. But, what does one mean by 'morphology' in this context? It
is only recently that a quantitative definition to this vague concept has been
proposed for scale-invariant structures.\ref\bbiii{R. Blumenfeld and R. C.
Ball, Phys. Rev. {\bf E 47}, 2298 (1993)} Note, however, that knowledge of the
distribution of the curvature along the surface is equivalent to knowledge of
the morphology of the asymptotic structure. For example, in the present context
a popular measurable quantity is the dimension of the growth that relates to
the time dependence of the size of the growth (the size can be defined by the
radius of gyration for an aggregate or by the equivalent circular
capacitor\ref\bbiv{R. C. Ball and R. Blumenfeld, Phys. Rev. {\bf A 44}, R828
(1991)}). This quantity relates directly to the third moment of the growth
probability distribution.\ref\hbb{T. C. Halsey, Phys. Rev. Lett. {\bf 59}, 2067
(1987); R. C. Ball and M. J. Blunt, Phys. Rev. {\bf A 39}, 6545 (1989)} A
customary generalization has been to study higher moments of this distribution.
These can be cast in one function termed the multifractal function (or
spectrum). Although this author believes that the multifractal function lacks
sensitivity for useful characterization of the surface's morphology for these
growth processes, it is nevertheless a signature of the structure. Therefore
one acid test of the present theory is whether it can lead to prediction of the
fractal dimension in particular and this entire function in general. In fact,
the observation in section 5 that the curvature entertains a power-law
distribution even for well-behaved distributions of the QP's already points to
the origin of scale invariance and fractality. Thus this formalism has the
potential to {\it derive} the onset of such behaviour (rather than assume scale
invariance, as is usually done in the literature for analytical calculations).
Indeed, full knowledge of the curvature statistics is sufficient to determine
these quantities as follows: The local curvature of the evolving surface is
proportional to the local field gradient normal to the surface. The latter can
be related in these processes, through some constitutive relation, to the
growth probability, $p$. For example, a popular such relation is
\eqn\Ei{p = \left|\nabla\Phi\right|^\eta \sim K^\eta \ ,}
with $\eta$ a parameter that can be adjusted according to the system under
consideration. Thus from the knoweldge of $\cP_1(K)$ it should not be too
difficult to derive the distribution of $p$ and therefore the entire
multifractal function (see below).

To facilitate such a calculation, the next stage consists of employing the
existence of Liouville's theorem due to the Hamiltonian description and
considering the spatial distribution of the QP. This was carried out in two
ways:

(i) First, considering an ensemble of initial conditions one defines a measure,
$\mu(H)$, from which the partition function, $\cZ$, can be calculated. Writing
down the explicit dependence of the curvature along the surface on the
locations of the QP one can then calculate directly the moments of the
probability density of the curvature. As in the case of Gibbs measure, $\mu(H)
= e^{-\beta H}$, one can identify a Lagrange multiplier, $\beta$, that
corresponds to 'smearing' of the trajectory due to fluctuations in the process.
This introduction of noise is analogous to (but seems to this author somewhat
more natural than) introducing noise effects directly in the EOM of the QP.
Nevertheless, this equivalence can be probably demonstrated by an analogue of
the usual fluctuation-dissipation theorem, an exercise that has not been
attempted in this presentation.

(ii) The second is a more dynamical approach: The existence of a Hamiltonian
implies by Liouville's theorem that the volume occupied by the system in the
4N-dimensional phase space is incompressible. Combining this with the
observation that the statistics of the surface flows towards a stable fixed
point, immediately leads to the simplified master equation, $\Dviii$, for the
distribution of the locations of the QP in the mathematical plane. I have not
attempted here to solve this equation in any limit or approximation. Rather, I
showed that the solution to that master equation yields all the needed
information on the asymptotic morphology of the evolving surface, again by
using the explicit dependence of the curvature on the locations of the QP.

Thus the curvatures statistics can be calculated in the present approach and
therefore the entire multifractal function as well as other relevant
quantities. For example, combining the approximation $\Dvv$ with the
constitutive relation $\Ei$, one can relate the growth probability
distribution, $\cP_2(p)$ to the distribution of the QP as follows:
\eqn\Ei{\cP_2(p) = Const. p^{-1-{1\over {2\eta}}} \cP_0 \left[1 - (2 C)^{1/2}
p^{-{1\over{2\eta}}}\right]\ .}
Again, the power-law decay of this probability, which appears {\it regardless}
of the behaviour of $\cP_1$ points towards the possible origin of fractality in
the system.

Thus, to the best of this author's knowledge this theory represents currently
the only approach that can lead to a quantitative calculation of all these
properties from first-principles.

\vfill
\eject
\appendix{A}{An integrable case study: N-symmetric growth with arbitrary
initial conditions}
\bigskip

In this Appendix I consider a particular class of arbitrary initial conditions
for which the system is integrable and the solution to the set of EOM can be
obtained explicitly. Let us consider an initial surface that is represented at
$t=0$ by the form
\eqn\Bi{\gamma(s,0) = e^{is} + \sum_{n=1}^N R_n\ln\bigl(e^{is}-P_n(0)\bigr)\ ,}
where $R_n = \prod_{m'\neq n}^N \bigl[\bigl(P_n(0) - Z_{m'}(0)\bigr)/
\bigl(P_n(0)-P_{m'}(0)\bigr)\bigr]$. It can be shown that the
surface-tension-free propagating curve, $\gamma(s,t)$, can also be described by
this form for any later time if the values of $P_n$ and $Z_n$ are substituted
by their time-dependent values. The form $\Bi$ is valid for any number of QP,
where $R_n$ should be interpreted as the residues of the function $F'=dF/dz$
when a contour-integral is taken around the $n$th pole, $P_n$. Moreover, one
can easily convince oneself from the EOM\sb that the number of QP, $N$, of each
kind is invariant under the EOM. Thus the growth problem consists now of
investigating the dynamics of $N$ zeros at $Z_n(t)=Z(t)e^{in\alpha}$, where
$\alpha\equiv 2\pi/N$, and $N$ poles at $P_n(t)=P(t)e^{in\alpha}$ ($Z(t)$ and
$P(t)$ are real functions of time). The initial values, $Z(t=0)$ and $P(t=0)$,
are completely aribitrary, as long as $P(0),\ Z(0) \neq 0$ and $P(0) \neq
Z(0)$. The dynamics of these singularities can be found by substituting
directly into the Eqs. $\Aiii$. An observation that is worthwhile to note is
that both from symmetry arguments and from direct analysis of the EOM one can
see that the motion of all the QP will be purely radial. Therefore, the
arguments $n\alpha$ stay constant and the only time-dependent variables are the
radial locations from the origin, $Z(t)$ and $P(t)$. Next we observe that
requiring that $F$ is holomorphic outside $\gamma$ implies that there is a sum
rule imosed on the locations of the singularities:\bbi$^,$\bii
\eqn\Biii{\sum_{n=1}^N Z_n(t) = \sum_{n=1}^N P_n(t)\ ,}
which can be shown to be identical to requiring that
$$\sum_{n=1}^N R_n = 0\ .$$

These constraints simplify the EOM that can now be written as
\eqn\Biiia{\eqalign{
-{1\over {Nx}} {\dot x} &= y/x - \cK \bigl[1 - 2/N + (1 + 2/N)y/x \bigr] \cr
-{1\over {Ny}} {\dot y} &= y/x - \cK (1 + y/x) \ ,\cr} }
where I have defined $x\equiv Z^N ,\ y\equiv P^N$ and $\cK\equiv
(1-xy)/(1-x^2)$. This system of equations displays a qualitatively different
behaviour when $P(0)$ is smaller or larger than $Z(0)$. In the first case cusp
singularities appear at a finite time which corresponds to the time that a zero
arrives at the UC. In the second case such singularities are avoided for any
finite time. The issue of particular cuspless growth solutions due to
specialized choices of the initial conditions has received some attention in
the literature.\ref\kad{S.D. Howison, SIAM J. Appl. Math. {\bf 46}, 20 (1986);
D. Bensimon and P. Pelce, Phys. Rev. {\bf A 33}, 4477 (1986); M.
Mineev-Weinstein and S.P. Dawson, preprint (1993)} This issue is not directly
relevant to the main thrust of this paper, but it should be pointed out that
recent calculations\ref\biii{R. Blumenfeld, unpublished} for the present
N-symmetric case suggest that the non-cuspiodal solution for $P(0)>Z(0)$ is
unstable for small perturbations in $\alpha$, under which it flows into a
cusp-forming solution.

To emphasize a point made in the text I intentionally choose initial conditions
that lead to cusp formation, $P(0)<Z(0)$ and arg$(P_n)={\rm arg}(Z_n)$. The
point is that regardless of whether the solution breaks down after a finite
time or not, the dynamics is Hamiltonian up to the moment of breakdown. In
other words, the dynamics remain Hamiltonian (and for the present case,
integrable) as long as the EOM are valid. When cusps form the very EOM cease to
be valid and the entire framework of transforming the growth problem into the
many-body dynamics no longer holds. Indeed, one of the goals in the text is to
extend the formalism with the aid of surface energy so as to make it hold for
$t\to\infty$ {\it regardless of initial conditions}. Since this has been
achieved in the text it makes no difference to the present analysis whether the
description of the surface-tension-free curve holds for a finite or infinite
time.

Inspection of the EOM shows that if the zero and the pole meet, say at $r_0$,
they keep moving together at an exponential rate,
$$r(t) = r_0 e^{Nt}\ .$$
This observation, however, is academic because once a pole and a zero meet they
'anihilate'. This can be verified by inspecting the form of the conformal map
$\Aii$: upon encounter $P_n=Z_n$ and therefore the corresponding terms
$(z-Z_n)/(z-P_n)$ cancel, and these QP disappear from the scene.

Using $\Bi$ and a straightforward manipulation yields for the explicit form of
the interface at any time $t>0$:
\eqn\Biiiz{\gamma(s,t) = e^{is} - {{Z(t) (1 - y/x)}\over N} \sum_{n=1}^N
e^{2n\pi i/N} \ln\bigl(e^{is} - P(t) e^{2n\pi i/N}\bigr)\ .}

Carrying out a rather tedious manipulation of the EOM $\Biiia$, one can show
that the following is a constant of the motion
\eqn\Biiic{{{y^{1-2/N}}\over{x-y}} + L(y) = Const\ ,}
where
\eqn\Biiid{L(y) \equiv {1\over N} \int^{y^2} {{u^{-1/N} du}\over{1-u}}\ .}
The EOM can be integrated out now and the trajectories of $P$ and $Z$ can be
found explicitly.

\smallskip
We can now find the canonical action-angle variables in terms of the original
coordinates. The action variable can be immediately set to
\eqn\Biiie{J = {{y^{1-2/N}}\over{x-y}} + L(y)\ ,}
which we know is a constant of the motion. The Hamiltonian is then
$$H = \omega J\ ,$$
with $\omega$ some constant frequency and with the angle variable $\Theta =
\omega t + \Theta_0$. The fact that we have only one action and one angle
variables reflects the degeneracy of the problem due to the $N$-rotational
symmetry, where only $Z(t)$ and $P(t)$ remain the relevant degrees of freedom.
Thus we have an integrable Hamiltonian that depends only on half the number of
degrees of freedom, $J$. Substituting $x$ from Eq. $\Biiic$ or $\Biiie$ into
the second equation of the set $\Biiia$ yields immediately the result for
$y(t)$ in the form
\eqn\Biiif{t - t_0 = {1\over N} \int^y {{1 - \xi^2(1 + \cL)^2}\over{\xi (1 -
\xi^2)}} d\xi \ \ \ ;\ \ \ \cL(y)\equiv {{y^{-2/N}}\over{J - L(y)}}\ .}
And substituting this into the first of the equations $\Biiia$ gives the
corresponding solution for $x(t)$.

As mentioned in the text, in this treatment, the prefactor in front of the
conformal map $F$, $A(t)$, is taken to be unity. Since an implicit assumption
in the general formulation of the problem is that the flux into the growth is
constant in time then the total area enclosed by the surface should increase
linearly with time. Thus by maintaining a unity prefactor the growth is in fact
rescaled at each time step by $1/A(t)$. For this reason a plot of this surface
will reveal sections of the boundary that seem to retreat with time, although
the actual physical surface always grows outwards. The evolution of this
prefactor follows a first order ODE as has already been discussed in the text.

\bigskip
{\bf Acknowledgement}

I thank R. C. Ball, M. Mineev-Weinstein, G. Berman and D. D. Holm for fruitful
discussions and helpful comments. I am grateful to A. R. Bishop for critically
reading the manuscript and for comments.

\bigskip

\centerline {\bf FIGURE CAPTIONS}

\item {1.} The evolution of the rescaled surface discussed in the text and in
Appendix A for $N=3$. a) The unregularized surface with $\sigma=0$ at times
t=0. ($\diamondsuit$), 0.11, 0.22 and $0.267$. The surface develops cusps at
t=0.2673. b) The regularized surface shown at times t=0. ($\diamondsuit$), 3.0,
6.0 and 9.0. After rescaling the last three curves are indistinguishable and
indicate a uniform growth.

\bigskip

\vfill\eject\immediate\closeout\rfile
\baselineskip=18pt\centerline{{\bf REFERENCES}}\bigskip\frenchspacing%
\input refs.tmp\vfill\eject\nonfrenchspacing
\bye